\documentclass[aps,prl,floatfix,twocolumn,reprint,amsmath,amssymb,superscriptaddress,UTF8]{revtex4-1}
\usepackage{amsfonts}
\usepackage{mathrsfs}
\usepackage{amsmath}
\usepackage{color}
\usepackage{natbib}
\usepackage{graphicx}
\usepackage{bm}
\usepackage{amssymb}
\usepackage{xspace}
\usepackage{epstopdf}
\usepackage{dcolumn}
\usepackage{longtable}
\usepackage{array}
\usepackage{makecell}
\usepackage{tabulary}
\usepackage{multirow}
\newcolumntype{C}[1]{>{\centering\arraybackslash}p{#1}}
\newcolumntype{L}[1]{>{\flushleft\arraybackslash}p{#1}}

\usepackage{multirow}
\usepackage{epsfig}
\usepackage[normalem]{ulem}
\usepackage{amsmath}
\usepackage{braket}
\usepackage{bbold}
\usepackage{natbib}

\usepackage[colorlinks=true, letterpaper=true, pdfstartview=FitV, linkcolor=blue, citecolor=blue, urlcolor=blue]{hyperref}

\makeatletter

\newcommand{\Rmnum}[1]{\expandafter\@slowromancap\romannumeral #1@}
\makeatother

\begin{document}

\title{Predictable gate-field control of spin in altermagnets with spin-layer coupling}

\author{Run-Wu Zhang}
\thanks{These authors contributed equally to this work.}
\affiliation{Key Lab of advanced optoelectronic quantum architecture and measurement (MOE), Beijing Key Lab of Nanophotonics $\&$ Ultrafine Optoelectronic Systems, and School of Physics, Beijing Institute of Technology, Beijing 100081, China}

\author{Chaoxi Cui}
\thanks{These authors contributed equally to this work.}
\affiliation{Key Lab of advanced optoelectronic quantum architecture and measurement (MOE), Beijing Key Lab of Nanophotonics $\&$ Ultrafine Optoelectronic Systems, and School of Physics, Beijing Institute of Technology, Beijing 100081, China}

\author{Runze Li}
\affiliation{Key Lab of advanced optoelectronic quantum architecture and measurement (MOE), Beijing Key Lab of Nanophotonics $\&$ Ultrafine Optoelectronic Systems, and School of Physics, Beijing Institute of Technology, Beijing 100081, China}

\author{Jingyi Duan}
\affiliation{Key Lab of advanced optoelectronic quantum architecture and measurement (MOE), Beijing Key Lab of Nanophotonics $\&$ Ultrafine Optoelectronic Systems, and School of Physics, Beijing Institute of Technology, Beijing 100081, China}

\author{Lei Li}
\affiliation{Key Lab of advanced optoelectronic quantum architecture and measurement (MOE), Beijing Key Lab of Nanophotonics $\&$ Ultrafine Optoelectronic Systems, and School of Physics, Beijing Institute of Technology, Beijing 100081, China}

\author{Zhi-Ming Yu}
\email{zhiming\_yu@bit.edu.cn}
\affiliation{Key Lab of advanced optoelectronic quantum architecture and measurement (MOE), Beijing Key Lab of Nanophotonics $\&$ Ultrafine Optoelectronic Systems, and School of Physics, Beijing Institute of Technology, Beijing 100081, China}

\author{Yugui Yao}
\email{ygyao@bit.edu.cn}
\affiliation{Key Lab of advanced optoelectronic quantum architecture and measurement (MOE), Beijing Key Lab of Nanophotonics $\&$ Ultrafine Optoelectronic Systems, and School of Physics, Beijing Institute of Technology, Beijing 100081, China}
\date{\today}
\begin{abstract}
Spintronics, a technology harnessing  electron spin for information transmission, offers a promising avenue to surpass the limitations of conventional electronic devices. While the spin directly interacts with the magnetic field, its control through the electric field is generally more practical, and has become a focal point in the field of spintronics. Current methodologies for generating spin polarization via an electric field generally necessitate spin-orbit coupling. Here, we propose an innovative mechanism that accomplishes this task without dependence on spin-orbit coupling.
Our method employs two-dimensional altermagnets with valley-mediated spin-layer coupling (SLC), in which electronic states display symmetry-protected and valley-contrasted spin and layer polarization. The SLC facilitates predictable, continuous, and reversible control of spin polarization using a gate electric field. Through symmetry analysis and \emph{ab initio} calculations, we pinpoint high-quality material candidates that exhibit SLC. We ascertain that applying a gate field of $0.2$ eV/\AA~ to  monolayer Ca(CoN)$_2$ can induce significant spin splitting up to 123 meV.
As a result, perfect and switchable spin/valley-currents, and substantial tunneling magnetoresistance can be achieved in these materials using only a gate field. These findings provide  new opportunities for generating predictable spin polarization and designing novel spintronic devices based on coupled spin, valley and layer physics.
\end{abstract}
\maketitle

\textit{\textcolor{blue}{Introduction.}}--
The discovery of various spin-dependent transport phenomena~\cite{julliere1975tunneling, baibich1988giant, binasch1989enhanced, miyazaki1995giant, moodera1995large, berger1996emission, milner1996spin, slonczewski1996current, tsymbal2003spin, zhang2004roles, merodio2014spin, jungwirth2014spin, wadley2016electrical, bodnar2018writing, vzelezny2018spin, shao2021spin} has led to the emergence of the field of spintronics, which has generated extensive interest in both fundamental research and application design.
Two prior conditions for spintronics~\cite{vzutic2004spintronics, awschalom2007challenges, fert2008nobel, bader2010spintronics, dieny2020opportunities, kim2022ferrimagnetic, han2023coherent} are the active control of spin degrees of freedom and the efficient generation of spin polarization.
As spin is a type of angular momentum, magnetic field represents the most natural and intuitive way to control it.
Specifically, under a magnetic field $\boldsymbol{B}$, the energy level of an electron with spin angular momentum $\boldsymbol{s}$ will be shifted by
$\Delta\propto -\boldsymbol{B}\cdot  \boldsymbol{s}$, a phenomenon known as the Zeeman effect~\cite{wolf2001spintronics, schapers2016semiconductor}.
However, magnetic control of spin is not favored by advanced electronic devices.


In practice, the most desirable approach for controlling spin polarization is static and electric control~\cite{ohno2000electric, sahoo2005electric, rovillain2010electric, chun2012electric, gong2013new, heron2014deterministic, noel2020non, ingla2021electrical, gonzalez2021efficient}. However, electric fields do not directly couple to spin, necessitating a mediated effect. Current schemes mainly rely on the spin-orbit coupling (SOC) effect~\cite{manchon2015new}, which couples an electron's motion or momentum to its spin, enabling indirect interaction between the electric field and spin. Besides, in the view of symmetry, the addition of an electric field may reduce the symmetry of systems and then makes the spin-degenerate bands split~\cite{yuan2013zeeman, yuan2014generation, liu2019band, zhao2022zeeman}. All these schemes however fail to provide a simple picture like the Zeeman effect to  predict the  behavior of spin under an electric field. Additionally, these schemes require materials with strong SOC effects, imposing severe constraints on the search for viable material candidates~\cite{bailly1968energies, taguchi1983point, west2012native, edwards2006electronic}.

\begin{figure*}
\includegraphics[width=16 cm]{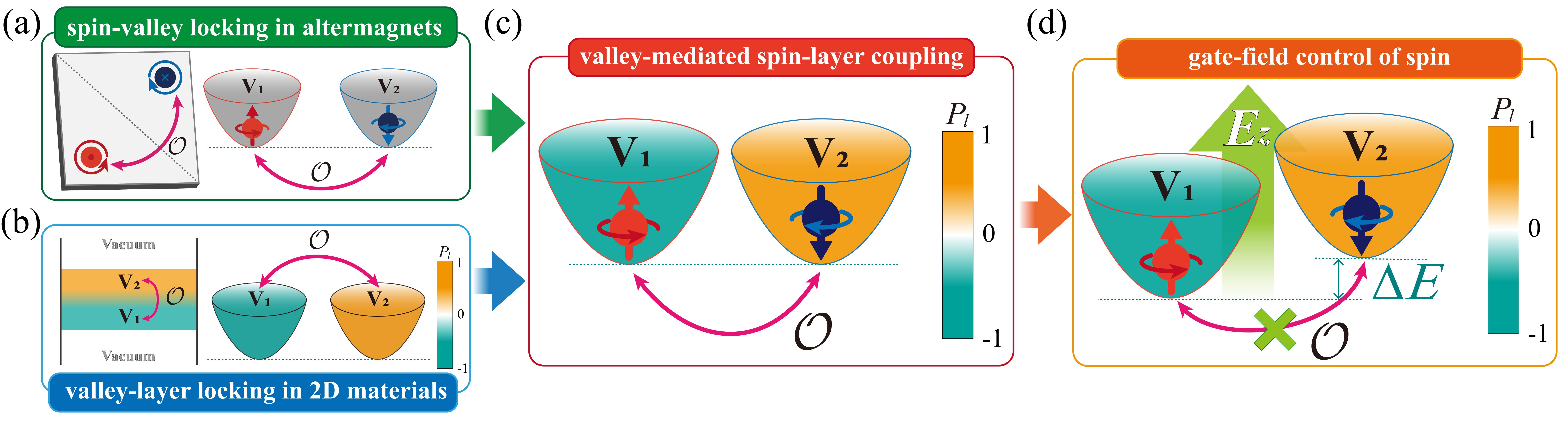}
\caption
{Illustration of the mechanism of valley-mediated SLC and  the electric  control  of spin.
(a) A generic altermagnet with two valleys $V_1$ and $V_2$ features intrinsic spin-valley  locking, which is protected by  certain (magnetic) crystalline symmetry $\mathcal{O}$ rather than time-reversal symmetry ${\cal{T}}$.
(b) Meanwhile,  a 2D valleytronic material may host $\mathcal{O}$-protected valley-layer locking, where the two valley states have opposite layer polarization ($P_l$).
(c) The combination of  spin-valley   and  valley-layer locking  leads to a  novel spin-valley-layer coupling: valley-mediated SLC in 2D altermagnets.
(d) This effect enables an intuitive, predictable and precise control of the spin polarization by electric gate field.}
\label{fig1}
\end{figure*}

Two-dimensional (2D) systems~\cite{xu2013graphene, fiori2014electronics, tan2017recent, du2021engineering} offer new possibilities.
Since most of the 2D materials have a finite thickness, a gate electric field normal to the  materials ($E_z$) can create   a layer-dependent electrostatic potential.
Then if the electronic states with spin-up and spin-down are respectively  localized at the top and bottom layers of the system (see Fig. \ref{fig1}), a spin-layer coupling (SLC) emerges.
With SLC, one directly knows that a gate field $E_z$ can produce an energy shift of $\Delta\propto -E_z d$ between the opposite spins, where $d$ is the layer distance, resulting in a gate-field control of spin in these novel 2D systems.
Importantly, the SLC scheme not only provides an intuitive and predictable way to control the spin by electric method, but also does not require the SOC effect.


In this work, we first show that the  SLC effect can naturally appear in a novel kind of 2D altermagnets~\cite{vsmejkal2022emerging, vsmejkal2022giant, vsmejkal2022beyond, yuan2020giant, yuan2021prediction, hayami2019momentum, egorov2021antiferromagnetism, egorov2022spin, feng2022anomalous, liu2022spin, chen2022role} with valley-layer coupling, and then  detail the symmetry requirements for these systems.
Guided by the symmetry analysis,  we also identify a number of high-quality material candidates, where significant spin and valley polarization can be achieved by a moderate gate field.
At last, several interesting phenomena associated with SLC are revealed, such as a new design of giant tunneling magnetoresistance (TMR) device that does not require heterojunction structures,  spin-resolved interlayer exciton, and layer photogalvanic effect.
Thus, by presenting a disruptively new effect to realize predictable control of spin polarization by electric mean, our work resolves an outstanding challenge in the field of spintronics, and will facilitate future research on the design of spintronic devices.

\textit{\textcolor{blue}{Valley-mediated SLC.}}--
The SLC proposed here refers to a stable coupling between  spin and layer degrees of freedom of electronic  states in momentum space rather than real space, as it would make  electric-field control of spin polarization intuitive and predictable.
We begin by detailing the procedures that lead to SLC.
Since our focus is on generating spin polarization through electric means, the systems studied here should  break time-reversal symmetry  (${\cal{T}}$) but spin-neutral in the absence of  external fields. This suggests that the 2D systems are  antiferromagnetic or altermagnetic (AM) materials.
One might naively expect that interlayer antiferromagnetism can guarantee stable SLC. However, this is not always the case, as interlayer antiferromagnetism only ensures stable coupling between spin and layer in real space.
For example, the interlayer antiferromagnetism  with horizontal mirror can not have SLC.

{\renewcommand{\arraystretch}{1.4}
\begin{table*}
	\caption{List of the MLGs and the corresponding magnetic space groups (MSGs) that allow for the valley-mediated SLC.
In the third column, the possible location of the two valleys are presented. ($\bm k, \bm k^\prime$) denote two generic momentums in the BZ.
${\cal O}$ is the  symmetry operator connecting  $V_1$ and $V_2$. Here, we assume the  electronics states at $V_1$  ($V_2$) are either spin-up or spin-down.}
	\label{Table1}
	\begin{ruledtabular}
	\begin{tabular}{lllll}
			Lattice & \makebox[2cm][c]{BZ}  & $\left(V_1,\ V_2\right)$ & ${\cal O}$& MLGs (MSGs)   \tabularnewline
			\hline
			\multirow{3}{*}{Oblique} &
			\makebox[2cm][c]{\multirow{3}{*}{\includegraphics[scale=0.3]{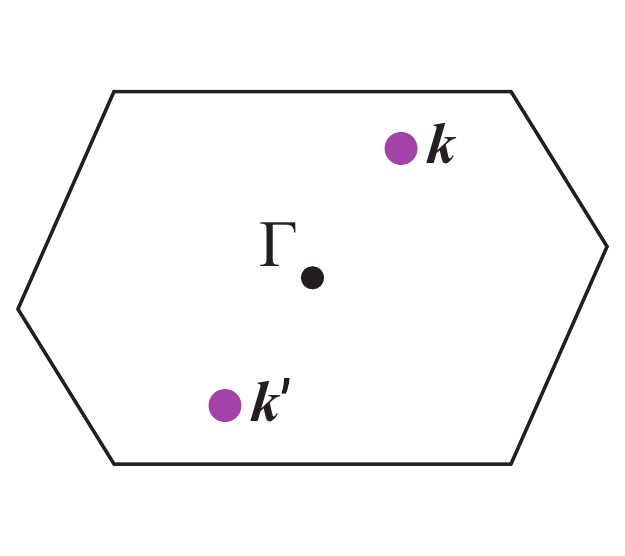}}}
			&\multirow{3}{*}{$\left(\bm k,\ \bm k'\right)$} &\multirow{3}{*}{$M_{z}\mathcal{T}$} &\multirow{3}{*}{$4.3.14$ $(6.3.27)$, $5.3.19$ $(7.3.34)$} \tabularnewline
			& &  &  & \tabularnewline
			& & & &\tabularnewline
			\multirow{3}{*}{Rectangular} &
			\multirow{3}{*}{\makebox[2cm][c]{\includegraphics[scale=0.3]{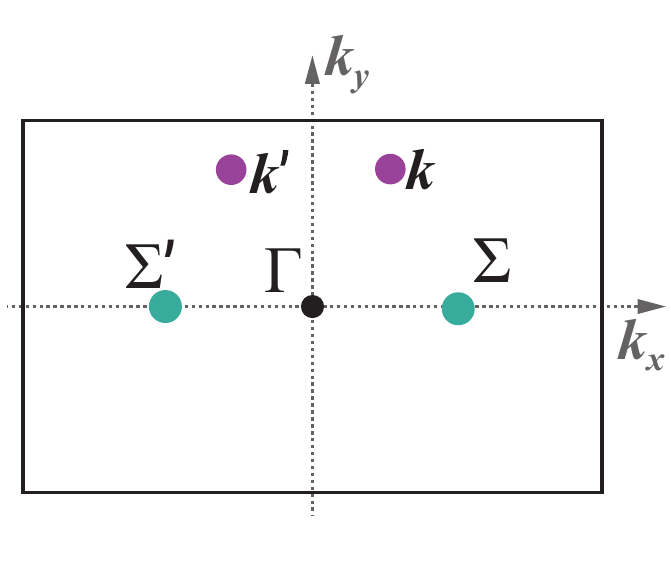}}} &\multirow{1}{*}{$\left(\bm k,\ \bm k'\right)$}  & $C_{2y}$ & $8.1.34$ $(3.1.8)$, $9.1.41$ $(4.1.15)$ \tabularnewline
			&  &$\left(\Sigma,\ \Sigma'\right)$ & $M_{z}\mathcal{T}$& $27.3.156$ $(25.4.158)$, $28.3.169$ $(26.5.172)$, $29.3.176$ $(26.5.172)$, $30.3.183$ $(27.4.181)$,  \tabularnewline
			&  &  & & $31.3.190$ $(28.5.189)$, $32.3.199$ $(31.5.216)$, $33.3.204$ $(29.5.202)$, $34.3.209$ $(30.5.209)$ \tabularnewline
			\tabularnewline
			\multirow{3}{*}{\makecell[l]{Centered\\ Rectangular}  } &\makebox[2cm][c]{\multirow{3}{*}{\includegraphics[scale=0.3]{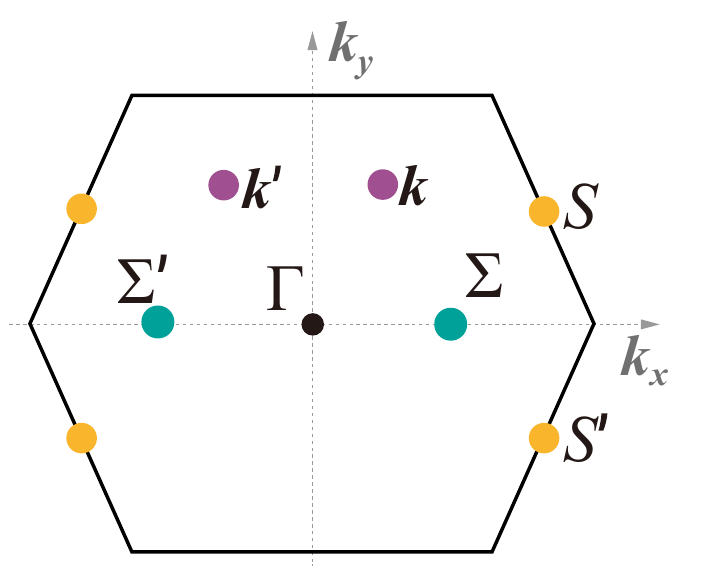}}} & \multirow{1}{*}{$\left(\bm k,\ \bm k'\right)$} & $C_{2y}$ & 10.1.45 (5.1.19) \tabularnewline
			& & \multirow{1}{*}{$\left(S,\ S'\right)$} & $C_{2x}$ & 22.1.122 (21.1.129) \tabularnewline
			& & \multirow{1}{*}{$\left(\Delta,\ \Delta'\right)$}& $M_{z}\mathcal{T}$ & 35.3.214 (38.5.269), 36.3.223 (39.5.282)\tabularnewline
			& & & &  \tabularnewline
			\multirow{3}{*}{Square} &\makebox[2cm][c]{\multirow{3}{*}{\includegraphics[scale=0.3]{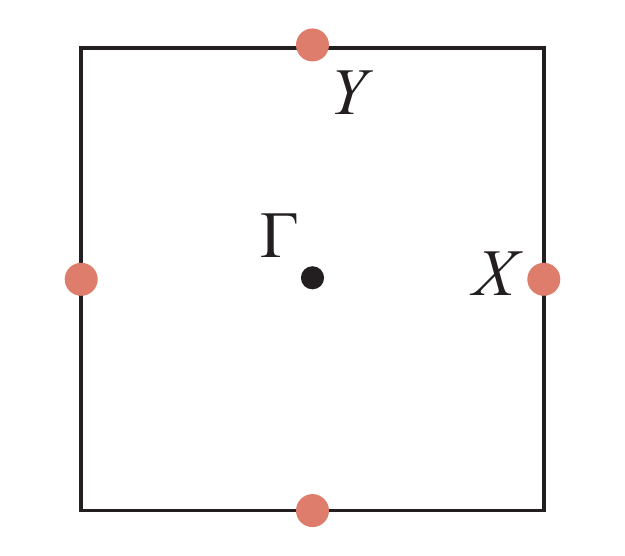}}} &\multirow{3}{*}{$\left(X, \ Y\right)$} & \multirow{3}{*}{$S_{4z}\mathcal{T}$} & \multirow{3}{*}{50.3.360 (81.3.695), 59.5.414 (115.3.943), 60.5.421 (117.3.960)} \tabularnewline
			& & & &  \tabularnewline
			& & & &  \tabularnewline
			& & & &  \tabularnewline
			\multirow{3}{*}{Hexagonal} &\makebox[2cm][c]{\multirow{3}{*}{\includegraphics[scale=0.3]{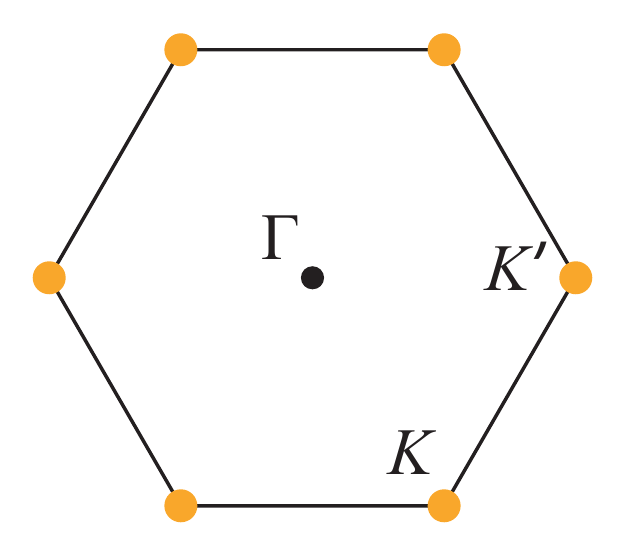}}} & \multirow{3}{*}{$\left(K,\ K'\right)$}  & $C^{\prime}_{21}$ & 67.1.467 (149.1.1251) \tabularnewline
			& & & \multirow{2}{*}{$M_{z}\mathcal{T}$} &  \multirow{2}{*}{74.3.494 (174.3.1365), 78.5.514 (187.3.1441), 79.5.519 (189.3.1453)}\tabularnewline
			& & & &  \tabularnewline
		\end{tabular}
	\end{ruledtabular}
\end{table*}


Instead, we  consider a generic AM material, where the  conduction or valence band edge (referred to as the ``valley") is not degenerate.
Due to the altermagnetism, the valleys must come in pairs with opposite spin polarization, leading to intrinsic spin-valley locking (see Fig. \ref{fig1}a).
Furthermore, recent research has demonstrated that strong valley-layer coupling can be achieved in certain 2D systems~\cite{yu2020valley}, where the electronic states in different valleys have strong but opposite layer polarization (see Fig. \ref{fig1}b).
By combining spin-valley and valley-layer locking (see Fig. \ref{fig1}c), the electronic states with opposite spin polarization (in different valleys) will exhibit opposite layer polarization.
Consequently, we obtain the desired SLC effect, which here is mediated by the valley degree of freedom.
Given the SLC effect, an uniform electric field can act as an uniform magnetic field, thereby enabling control the spin of the systems without the accompanying disadvantages of magnetic fields in practice (see Fig. \ref{fig1}d).

\textit{\textcolor{blue}{Symmetry requirements.}}--
We analyze the symmetry conditions for the valley-mediated SLC.
For clarity, we consider a  2D collinear AM system with only two valleys labeled as $V_1$ and $V_2$,  as shown in  Fig. \ref{fig1}.
The  N\'eel vector  of the  system points out of the plane.
We assume that  the $z$-axis is normal to the 2D AM system, and the  $z=0$ plane locates at the mid-plane of the system.
Then the system can be considered as  two layers: the top layer with $z>0$ and the bottom layer having $z<0$.
For each Bloch state $|\psi({\bm k})\rangle$, we can define a spin (layer) polarization $P_{s(l)}=\langle\psi({\bm k})|\hat{P}_{s(l)}|\psi({\bm k})\rangle$ with $\hat{P}_{s(l)}$ the spin (layer) operator. $P_{s}>0$ ($P_{s}<0$) and $P_{l}>0$ ($P_{l}<0$)  indicate that the state has more weight distributed in the spin-up (spin-down) and top (bottom) layer, respectively.
Therefore, to realize the valley-mediated SLC, the following symmetry requirements must be satisfied.

(i) Symmetries that guarantee vanishing  spin or layer polarization at $V_{1(2)}$  should be broken, such as  horizontal mirror ${\cal{M}}_z$, which  makes  $P_l=0$  for all the  electronic states. Besides, $V_{1(2)}$ can not have operators that reverse the spin and layer polarization.

(ii) We can divide the symmetries of the AM system  into two parts: $\cal{R}$ and $\cal{O}$.
 $V_1$ and $V_2$ are invariant (interchanged) under the  operators in $\cal{R}$ ($\cal{O}$), as follows:
\begin{equation}
\begin{aligned}
{\cal{R}}V_{1(2)}=V_{1(2)}, \ \ \ {\cal{O}}V_{1(2)}=V_{2(1)}.
\end{aligned}
\end{equation}
In order to achieve valley-contrasted spin and layer polarization, any operator in  $\cal{R}$  must maintain both spin and layer polarizations for each valley:
\begin{eqnarray}
  {\cal{R}}P_{l(s)}(V_{i}){\cal{R}}^{-1} &=&P_{l(s)}(V_{i}),
\end{eqnarray}
with $i=1,2$.
Meanwhile,  any operator in   $\cal{O}$  reverses the polarizations:
\begin{eqnarray}
  {\cal{O}}P_{l(s)}(V_{1/2}){\cal{O}}^{-1} &=& -P_{l(s)}(V_{2/1}).
\end{eqnarray}
Notice that all the external fields that break symmetry ${\cal{O}}$ can lead both valley and spin polarization. 
However, only  electric gate field can provide an intuitive and predictable control of the spin, as illustrated in Fig. \ref{fig1}d.

Using these two conditions, we conducted a search through all 528 magnetic layer groups (MLGs) to identify those that may host valley-mediated SLC. Our findings apply to systems both with and without SOC, and are summarized in Table \ref{Table1}. This table provides a systematic and specific guide for identifying material candidates.

\textit{\textcolor{blue}{High-quality material candidates.}}--
In addition to developing design principles, it is equivalently important to identify high-quality material candidates. Here, we demonstrate that the family of monolayer decorated transition metal nitrides $A$($B$N)$_2$ ($A=$ Mg, Ca, Zn and $B=$ Mn, Fe, Co) materials are the most likely candidates to achieve the valley-mediated SLC. Since all nine kinds of materials share  similar crystalline structures and electronic bands, we  mainly focus on Ca(CoN)$_2$ in the main text. Further details regarding the properties of the remaining material candidates can be found in the \textit{Supporting Materials} (SM).

\begin{figure}
\includegraphics[width=1.0\columnwidth]{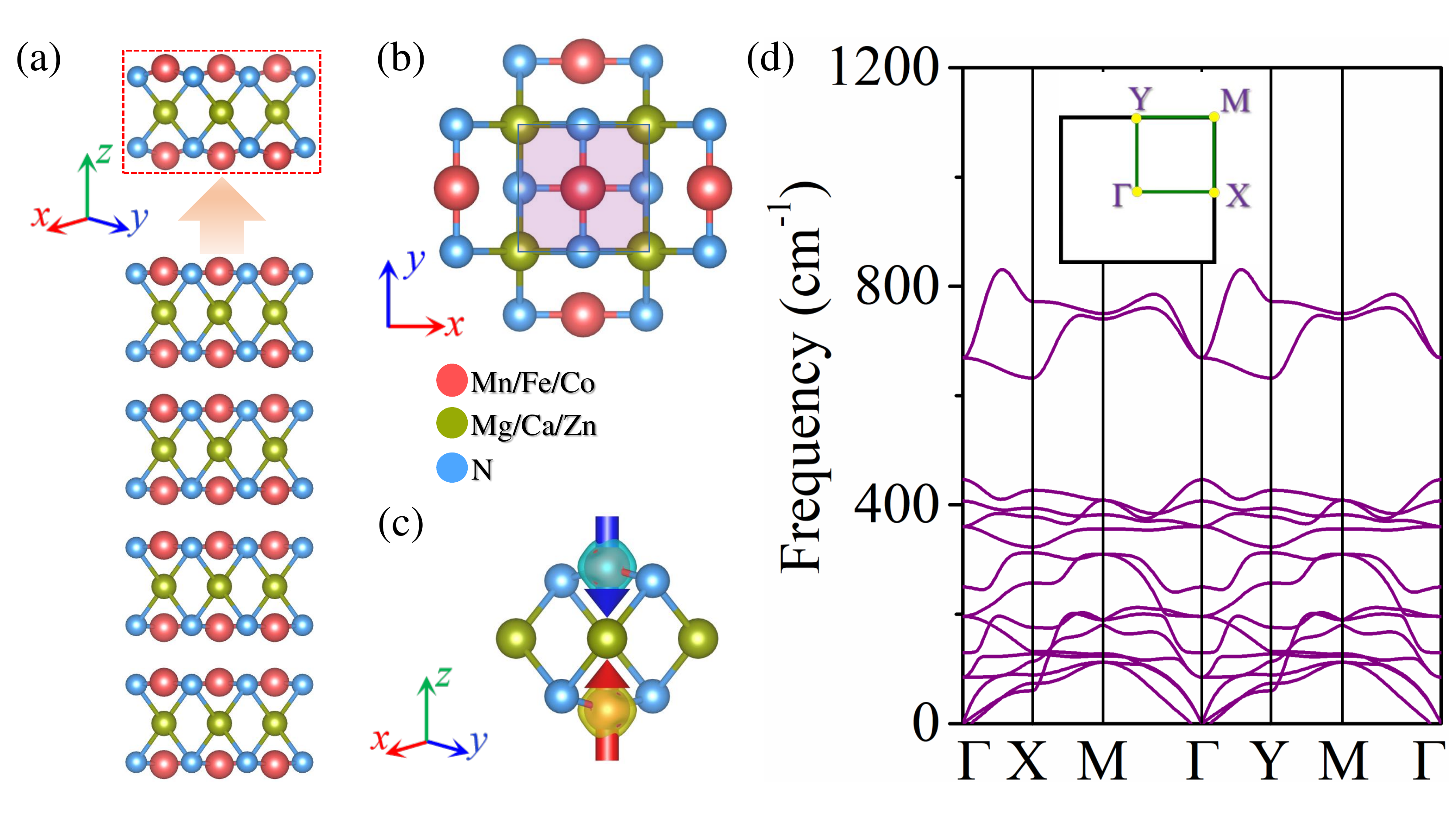}
\caption
{(a) Schematic showing the process of stripping monolayer candidates (red dashed area) from the bulk.
(b) Top view of the crystal structure of monolayer $A$($B$N)$_2$.  The light purple region denotes the primitive cell. (c) Spatial spin-density distribution of ML-CaCoN, showing the magnetic moments are mainly localized around the top and bottom Co atoms with opposite directions. (d) Phonon spectrum of the ML-CaCoN. The insert is  the BZ.}
\label{fig2}
\end{figure}

The 3D bulk Ca(CoN)$_2$ is a layered material that has been demonstrated to be energetically and dynamically stable, as well as easy to synthesize, peel, and transfer. Notably, the exfoliation energy for monolayer Ca(CoN)$_2$ (ML-CaCoN) is calculated to be $\sim$1.19 $J/m^{2}$, comparable to that of two representative 2D materials: Ca$_2$N ($\sim$1.09 $J/m^{2}$)~\cite{zhao2014obtaining} and HfGeTe ($\sim$0.98 $J/m^{2}$)~\cite{guan2017two}. Therefore, obtaining ML-CaCoN from its bulk materials through mechanical exfoliation is feasible. We also confirm that ML-CaCoN is dynamically stable (see Fig. \ref{fig2}d) and features thermal stability up to 300 K (see SM).

\begin{figure}
\includegraphics[width=1.0\columnwidth]{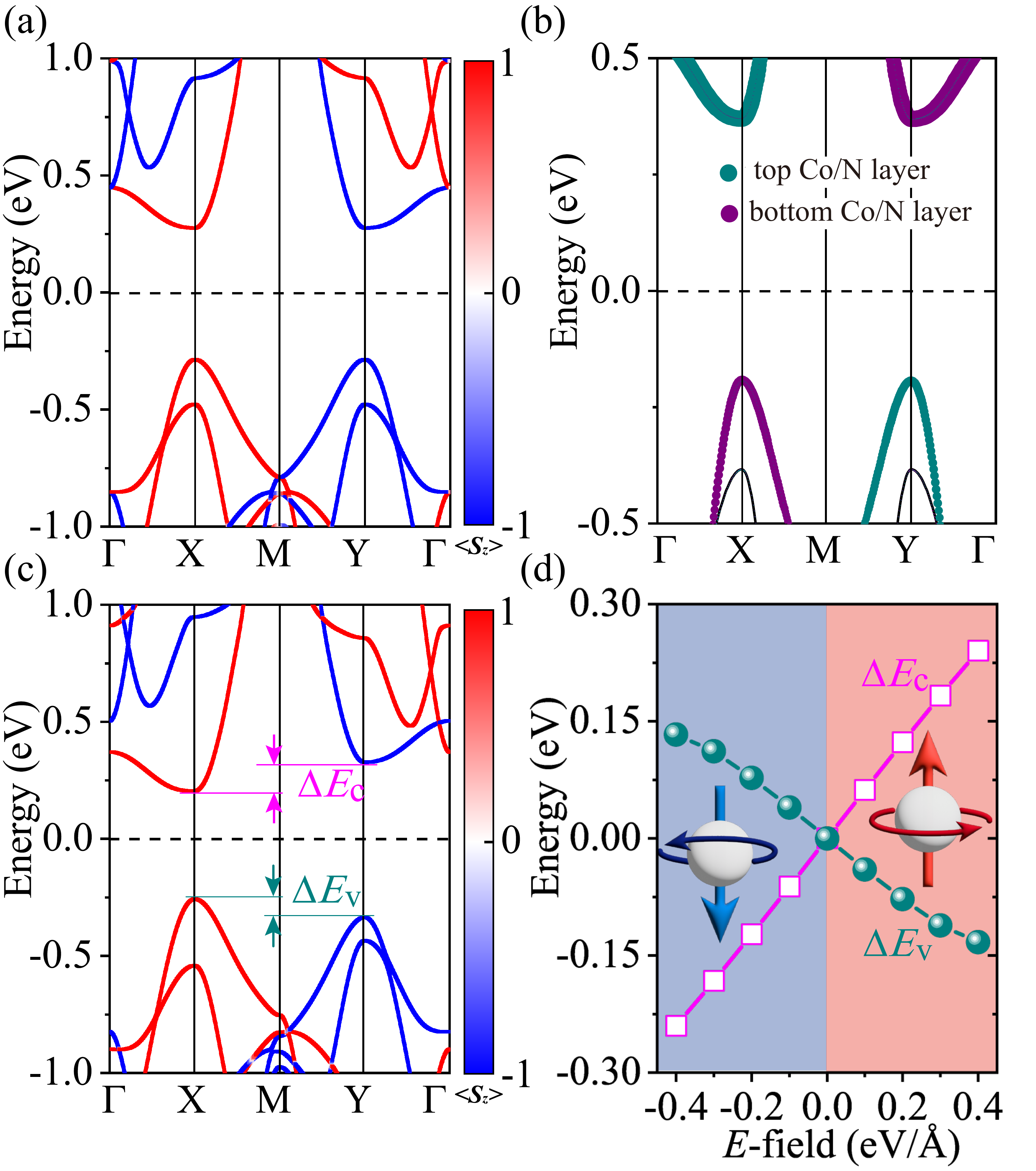}
\caption
{(a) Band structure of ML-CaCoN in the AM configuration with SOC. The color denotes the out-of-plane spin polarization ($s_z$). (b)  Orbital-projected band structures, showing the low-energy bands are mainly contributed by Co and N atoms. (c) Band structure of ML-CaCoN under a gate field of $E_z=0.2$ eV/\AA. (d) Spin  (valley) splitting for VBM ($\Delta E_{v}$) and CBM ($\Delta E_{c}$) [indicated in (c)] versus the applied gate field.}
\label{fig3}
\end{figure}

ML-CaCoN has a square lattice structure with optimized lattice constant $a=b=3.55$ \AA, consisting of five atomic layers in the sequence of Co-N-Ca-N-Co (see Fig. \ref{fig2}a).
The top (bottom) Co and N atoms are  almost in the  same plane.
Remarkably, the separation between the two Co (N) layers is about 3.9 \AA, which is  as large as the typical interlayer spacing of  van der Waals heterostructures.
Detailed calculations confirm that the AM configuration of ML-CaCoN is mostly energy stable (see Fig. \ref{fig2}c).
The magnetic moments are mainly on the Co sites with a magnitude of $\sim 3 \mu_B$ and the N\'eel vector is along the $z$ axis.
All these results show that the ground state of ML-CaCoN belongs to MLG 59.5.414, which is exactly an MLG candidate listed in Table \ref{Table1}.

According to our symmetry analysis (see Table \ref{Table1}), ML-CaCoN will exhibit valley-mediated SLC as long as it has two valleys at the $X$ and $Y$ points.
In fact, these two valleys are connected by $\mathcal{TS}_{4z}$ symmetry (with $\mathcal{S}_{4z}$ being a roto-inversion operator), which maps spin-up to spin-down, and interchanges the layer polarization of the two valleys.
Figure \ref{fig3}a presents the band structure of the AM order with the spin aligned along the $z$ direction under the SOC effect. It is observed that  there indeed exist two valleys for both the conduction and valence bands at the $X$ and $Y$ points in the Brillouin zone (BZ).
Therefore, we can conclude that the valley states must feature valley-contrasted spin and layer polarization.
This  prediction is confirmed by our first-principles calculations.
By analyzing the spin and atomic orbital projections of the valley states, we find that both the conduction and valence bands residing at the $X$ valley are completely composed of spin-up, while those at the $Y$ valley  only contain spin-down (see Fig. \ref{fig3}a).
Additionally, the conduction  band at $X$ ($Y$) valley is mainly distributed in the top (bottom) Co/N layer, whereas the valence band at $X$ ($Y$) valley is mainly distributed in the bottom (top) Co/N layer (see Fig. \ref{fig3}b).

\textit{\textcolor{blue}{Gate field control of spin splitting.}}--
We then study the behavior of the valley states of ML-CaCoN  under a gate  field along the $z$ direction $E_z$.
Due to SLC, this behavior  can be simply predicted  without complicated calculations.
Since the gate field creates a layer-dependent electrostatic potential $\propto e E_z d$  ($e$ denotes the electron charge), the  electronic bands with  layer polarization $P_l>0$ ($P_l<0$) will move up  (down).
Therefore, for the conduction band,  the gate field behaves as a static magnetic field along $-z$ direction,  pushing up the spin-up (at the $X$ valley) while pulling down the spin-down (at the $Y$ valley).
For the valence band, the spin splitting is opposite because the layer polarization of the spin (and valley) states are reversed (see Fig. \ref{fig3}b).
Notice that the layer distance of the top and bottom Co (N) layers  have an important impact  on the spin splitting, similar to the effective  \textit{g}-factor in Zeeman effect.

The calculated band structure of ML-CaCoN under a gate field of $E_z=0.2$ eV/\AA, which is achievable in experiments, is shown in Fig. \ref{fig3}c}.
The results are in line with our expectations.
The spin splitting of the conduction band minimum (CBM) and valence band maximum (VBM) respectively reach up to 123 meV and 78 meV (see Fig. \ref{fig3}c).
Remarkably, all the conduction (valence) bands around the $X$ ($Y$) valley feature almost the same energy shift as CBM (VBM).
Therefore, the gate field in ML-CaCoN  induces not only a large  but also an almost uniform effective static  magnetic  field.
To the best of our knowledge, such an effective static magnetic field has not been reported before and cannot generally be generated by schemes other than the SLC scheme.
Approximately, we can define an effective magnetic field
\begin{eqnarray}
B_{c(v)}^{\text{eff}}\equiv \frac{\Delta E_{c(v)}}{g_s s_z},
\end{eqnarray}
where $\Delta E_{c}$ ($\Delta E_{v}$) is the energy splitting for CBM (VBM), $g_s$ is the effective \textit{g}-factor which here is assumed to be 2, and $s_z=1/2$ is the spin momentum.
Accordingly,  one knows that for the conduction  (valence) band of the ML-CaCoN, a gate field of  $E_z=0.2$ eV/\AA \ leads to a significant magnetic field as large as $\sim 1.062\times10^{3}$ T  ($\sim 0.673\times10^{3}$ T). 

We also calculate the spin splitting as a function of the gate field, and the results are shown in Fig. \ref{fig3}d, which clearly shows that a  continuous,  wide-range, and switchable control of  spin (valley) polarization is  achieved.
When the gate field is not strong enough, we obtain a  linear  relationship between the induced  $B_{c(v)}^{\text{eff}}$ and  the gate field $E_z$,
\begin{eqnarray}
B_{c(v)}^{\text{eff}}=\chi_{c(v)} E_z,
\end{eqnarray}
where the coefficients are obtained as $\chi_{c}= 5.15\times10^{3}$ T\AA/V and $\chi_{v}=2.71\times10^{3}$ T\AA/V (see SM).

\textit{\textcolor{blue}{Effective model for valley-mediated SLC.}}--
To better understand the physics of the valley-mediated SLC,  we establish a low-energy effective Hamiltonian for the ML-CaCoN.
The  magnetic point group of the $X$ and $Y$ valleys are  $m'm'2$.
The band representation for the CBM  ($\varphi_{c}^{X}$) and VBM  ($\varphi_{v}^{X}$) at the $X$ valley are   $^2\overline{E}$ and $^1\overline{E}$ respectively, while that of the CBM   ($\varphi_{c}^{Y}$) and VBM  ($\varphi_{v}^{Y}$) at the $Y$ valley are $^1\overline{E}$ and $^2\overline{E}$.
Using the four states $\{\varphi_{c}^{X}, \varphi_{v}^{X},\varphi_{c}^{Y}, \varphi_{v}^{Y}\}$ as the basis, the ${\bm k\cdot \bm p}$ effective model can be expressed as:
\begin{eqnarray}\label{ham}
\mathcal{H}&=&\Lambda\tau_{z}+v_1(k_{x}\tau_{x}+k_{y}\tau_{y})+v_2(k_{x}\tau_{x}-k_{y}\tau_{y})\sigma_{z},
\end{eqnarray}
where $\Lambda=0.281$ is the half of the band gap, $v_{1}=1.167$ eV/\AA~ and $v_{2}=-0.109$ eV/\AA~ for ML-CaCoN. 
Both $\sigma$ and  $\tau$ are Pauli matrices that act on the valley space and the basis of $\{\varphi_{c}^{V}, \varphi_{v}^{V}\}$ with $V=X$ and $Y$, respectively.
Since the basis $\{\varphi_{c}^{V}, \varphi_{v}^{V}\}$  have opposite layer polarization, $\tau$ can also be regarded as acting on the  layer index space.
Similarly, $\sigma$ can also be considered to operate in the spin space, as the low-energy states at the $X$ ($Y$) valley is spin-up  (spin-down) electrons.
Therefore,  the  physics of the  SLC is solely encoded in  the last term of Eq. (\ref{ham}) because it pairs $\tau$ with $\sigma$.
With the concept of the effective magnetic field, the influence of the gate field on the low-energy bands of ML-CaCoN can be approximately   written as:
\begin{eqnarray}
\mathcal{H}_{E}&=& -s_z g_s \ \text{diag}(B_{c}^{\text{eff}}, -B_{v}^{\text{eff}}, B_{c}^{\text{eff}}, -B_{v}^{\text{eff}}).
\end{eqnarray}

\textit{\textcolor{blue}{Discussions.}}--
The multifaceted progression of altermagnetic spintronics~\cite{jungwirth2016antiferromagnetic, baltz2018antiferromagnetic, jungwirth2018multiple, vsmejkal2018topological, vsmejkal2022anomalous, vsmejkal2022emerging} has rendered altermagnets a compelling and highly practical research domain. Boasting unique advantages over traditional ferromagnets, such as the absence of stray fields, superior intrinsic precession frequency, and exceptional stability under magnetic fields, altermagnets are being hailed as the foundation for the next generation of high-performance devices. When SLC meets altermagnetism, it presents a great opportunity to explore more possibilities in AM materials, thereby achieving the effect of making 1 + 1 greater than 2.

\begin{figure}
\includegraphics[width=1.0\columnwidth]{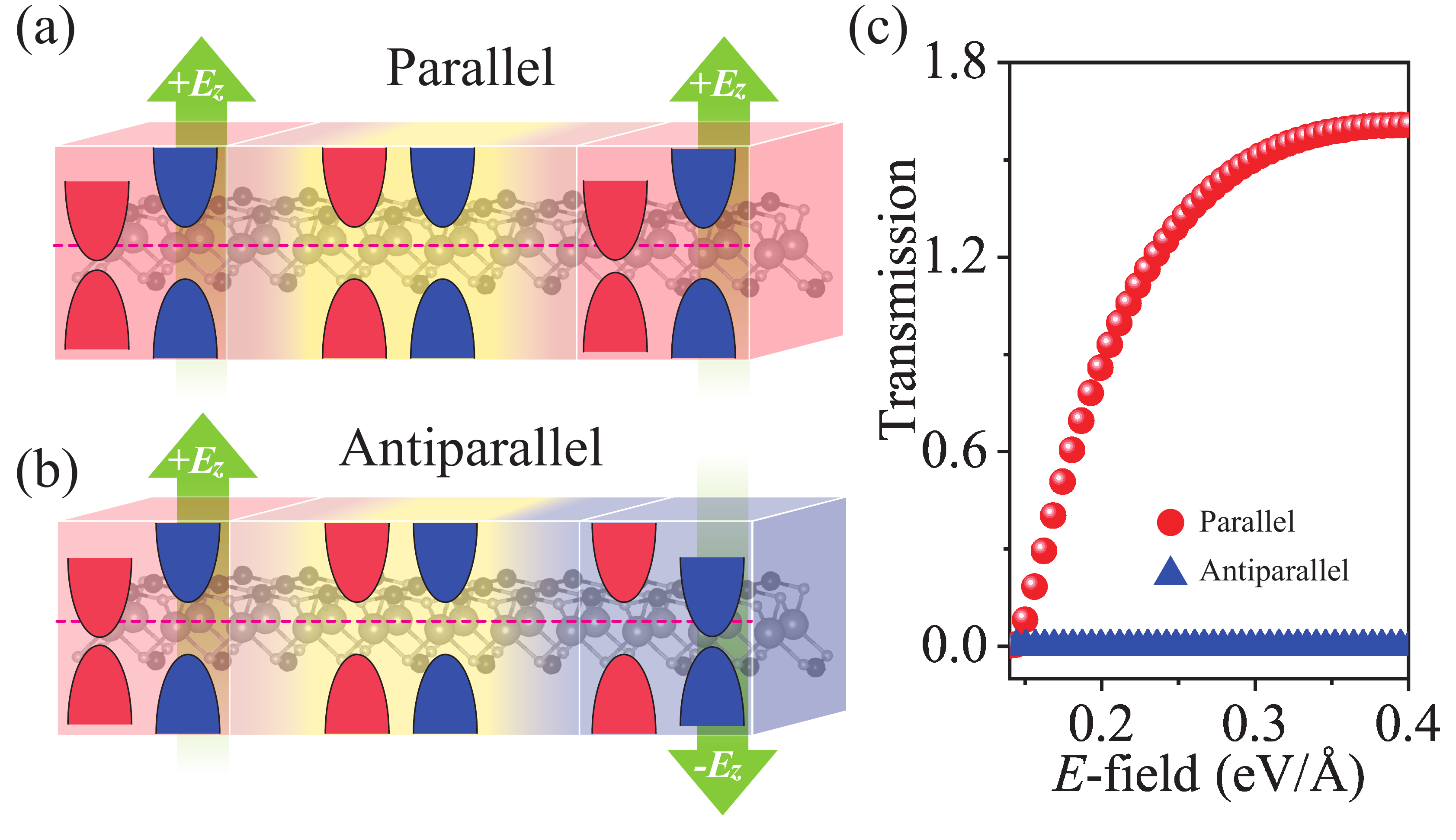}
\caption
{Schematic of giant TMR device based on ML-CaCoN. (a-b) Parallel and  antiparallel configurations of the TMR device, which are achieved  by applying gate field rather than magnetic field.
(c) Total transmission as a function of electric field ($E_z$) for the Ca(CoN)$_2$ tunnel junction in parallel (red dots) and antiparallel (blue triangles) configurations. Here, the Fermi level is set as 0.33 eV by doping electrons.}
\label{fig4}
\end{figure}

ML-CaCoN emerges as an optimal platform for generating a current that exhibits impeccable spin (valley) polarization. On doping the system with either electrons or holes and subjecting it to a gate field, the direction of spin (or valley) polarization can be effortlessly toggled by reversing the gate field's direction. This unique gate field induced switchability of spin polarization in ML-CaCoN hints at a novel design paradigm for achieving heightened tunneling magnetoresistance (TMR) using CaCoN alone. This approach significantly simplifies experimental procedures as it bypasses the need for the conventional conductive layer alternation between magnetic and non-magnetic materials, typically required in traditional TMR devices.
As depicted in Fig. \ref{fig4}a, in a parallel configuration (the Fermi level is raised to 0.33 eV by doping electrons), both the left and right terminals are subjected to positive electric fields (e.g. $E_z=0.2$ eV/\AA), simultaneously, the matching spin-polarized conduction channels between the electrodes culminate in an ultra-low-resistance state. In contrast, as demonstrated in Fig. \ref{fig4}b, in an anti-parallel configuration (for example, when the left terminal is subjected to $E_z=0.2$ eV/\AA, and the right to $E_z=-0.2$ eV/\AA), electrons moving from left to right must switch  spin, valley and layer indexes, namely changing from spin-up, $X$ valley and top layer to  spin-down, $Y$ valley and bottom layer. This  makes such transitions extremely less probable during transport. The overall transmission as a function of electric fields for the CaCoN tunnel junctions is obtained for both parallel (red dots) and antiparallel (blue triangles) configurations, as shown in Fig. \ref{fig4}c. Consequently, our design could potentially deliver a stronger suppression of antiparallel configurations, resulting in a more pronounced TMR effect compared to conventional TMR devices.

ML-CaCoN also presents a spectrum of intriguing and adjustable optical properties. For example, ML-CaCoN showcases valley- and spin-contrasted elliptical dichroism, in which left (or right) elliptically polarized light interacts with the inter-band transitions of the electrons possessing $X$ valley and spin-up index (or  $Y$ valley and spin-down index).
Moreover, the opposing layer polarization of valence and conduction bands at each valley leads to the formation of spin-resolved interlayer excitons ~\cite{yu2020valley} and layer photogalvanic effects under the illumination of left or right elliptically polarized light ~\cite{gao2020tunable}. This phenomenon results from the localization of optically excited electrons and holes within different layers of the ML-CaCoN structure.
Significantly, the characteristics of these excitons and photogalvanic effects can be modulated through the  gate field, as the gate field impacts the local band gap at the $X$ (or $Y$) valley.

\textit{\textcolor{blue}{Acknowledgments.}}
The work is supported by the National Key R\&D Program of China (Grant No. 2020YFA0308800), the National Natural Science Foundation of China (Grant No. 12061131002),  the Strategic Priority Research Program of Chinese Academy of Sciences (Grant No. XDB30000000),  the  National Natural Science Fund for Excellent Young Scientists Fund Program (Overseas) and the Beijing Institute of Technology Research Fund Program for Young Scholars.

\bibliography{ref}



\end{document}